# Validation of a Real-Time Manual Wheelchair Simulator through Biomechanical and Perceptual Measures: A Comparison with Overground Propulsion

Ateayeh Bayat[a,b*], Félix Chénier[c,b]

[a] Department of Systems Engineering, École de Technologie Supérieure (ÉTS), Montreal, Canada, [b] Centre for Interdisciplinary Research in Rehabilitation of Greater Montreal (CRIR), Montreal, Canada, [c] Department of Physical Activity Sciences, Université du Québec à Montréal (UQAM), Montreal, Canada

## Abstract

**Purpose:** Manual wheelchair simulators can provide a safe and controlled environment for propulsion training, biofeedback, and biomechanical assessment. However, their ecological validity must be established to ensure that simulator-based outcomes reflect overground propulsion. This study aimed to evaluate the ecological validity of a real-time manual wheelchair simulator by comparing biomechanical and perceptual outcomes during matched overground and simulator tasks.
**Materials and Methods:** Thirteen participants (10 able-bodied individuals, 3 experienced manual wheelchair users), performed propulsion tasks overground and on the simulator. Task segments included straight-line propulsion, acceleration, turning, ascending, and descending. Biomechanical outcomes were collected using instrumented wheels and compared between environments using temporal, kinetic, and waveform-based measures. Perceived realism and user experience were assessed using task-specific realism ratings, a post-test questionnaire, and semi-structured interviews.
**Results:** Temporal variables showed relatively small differences between overground and simulator propulsion, whereas kinetic variables showed larger discrepancies, consistent with previous simulator comparisons. Waveform patterns for force, moment, and power demonstrated high similarity between environments. Participants generally reported positive perceptions of realism, safety, and satisfaction, and highlighted the value of practicing wheelchair skills in a controlled environment.
**Conclusion:** The simulator demonstrated partial ecological validity, particularly for temporal and waveform-based propulsion outcomes. Its ability to accommodate the user's own wheelchair configuration may help preserve ergonomics and enhance realism, supporting its potential use as an assistive technology tool for manual wheelchair propulsion training.

**Keywords:** manual wheelchair; wheelchair simulator; assistive technology; ecological validity; propulsion biomechanics; virtual reality.

**Implications for rehabilitation**

- The proposed simulator may complement manual wheelchair training by allowing users, particularly novice or anxious users, to practice complex propulsion situations involving slopes, turns, and static or moving obstacles in a safe and controlled environment.
- The preservation of temporal propulsion patterns and waveform similarity supports the potential use of the simulator for propulsion training and future real-time biofeedback applications.
- Allowing users to train with their own wheelchairs and wheels may improve comfort, preserve individualized wheelchair ergonomics, and enhance perceived realism during simulator-based rehabilitation.

---

**CONTACT** Ateayeh Bayat ateayeh.bayat.1@ens.etsmtl.ca.

## Introduction

Manual wheelchair (MW) propulsion is a complex, multi-task activity that requires coordinated motor control and attention (Arlati et al., 2020; Nguyen et al., 2019). It is also a repetitive, high-load task that engages multiple muscles and joints (Qi et al., 2019) and can contribute to upper-limb injuries when propulsion biomechanics are unfavourable (Boninger, Koontz, et al., 2004). Shoulder and wrist pain are highly prevalent among wheelchair users, affecting up to 70% of individuals and potentially limiting daily mobility (Boninger et al., 2002). These limitations may contribute to reduced physical activity and a sedentary lifestyle, leading to secondary health consequences such as weight gain and additional upper-limb injuries (Boninger et al., 1999).

Since propulsion biomechanics have been associated with injury risk and propulsion efficiency (Boninger et al., 2003; Boninger, Impink, et al., 2004; Boninger, Koontz, et al., 2004), studying propulsion and providing appropriate training may help reduce injury risk while improving propulsion efficiency and overall wheelchair skills (Morgan et al., 2017).

Overground propulsion represents the most realistic condition for measuring and monitoring propulsion kinetics and temporal characteristics during real-world tasks using instrumented wheels (Cowan et al., 2008). However, conducting controlled biomechanical experiments overground remains challenging due to environmental variability and limited control over propulsion conditions (de Klerk, Vegter, et al., 2020; Kwarciak et al., 2011; van der Woude et al., 2001). As a result, laboratory-based simulators have been developed to study MW propulsion under controlled and repeatable conditions (Arlati et al., 2020; MacGillivray et al., 2020).

A major limitation of current MW simulators is ecological validity, a component of external validity that reflects the extent to which results obtained in a controlled setting can generalize to real-life behaviour (Andrade, 2018). In this context, ecological validity depends on whether propulsion biomechanics and user strategies observed on a simulator reflect those measured overground.

One key factor influencing this is the simulator's ability to accurately model the mechanical characteristics of propulsion, including inertia and frictional forces (de Klerk, Vegter, et al., 2020), while being able to model curvilinear motion and capture the effects of body movement (Chenier et al., 2014). In addition, the simulator should be capable of reproducing different propulsion conditions such as ascending, descending, and lateral slopes, as users modify their propulsion strategies under varying environmental demands (Richter et al., 2007).

Another important factor is the ergonomics of the simulator. The physical characteristics of the device on which the user propels can directly influence propulsion biomechanics. For example, axle position and seat height affect parameters such as push angle (Boninger et al., 2000; Boninger, Koontz, et al., 2004), while wheel radius influences rolling resistance and handrim size affects the magnitude and application rate of force (van der Woude et al., 2001). Therefore, enabling the use of the user's own wheelchair and wheels is important for achieving realistic and valid simulation.

Finally, the ability of a simulator to provide realistic sensory feedback also contributes to ecological validity. To replicate overground propulsion perception, simulators should provide appropriate visual, auditory, haptic, and vestibular feedback (Arlati et al., 2020; de Klerk, Vegter, et al., 2020).

In our previous work, we presented a real-time MW simulator (Bayat & Chénier, 2025) integrating a force sensor to capture forces applied to the wheelchair system, including inertial effects due to body motion, to improve dynamic modelling accuracy. The simulator also incorporates virtual reality and a motion platform, and is capable of modelling curvilinear motion and ramp conditions. Importantly, users can utilize their own wheelchairs and wheels, preserving the ergonomics of the system.

The objective of the present study was to evaluate the ecological validity of the simulator by comparing biomechanical outcomes between matched overground and simulator propulsion tasks and by assessing perceived realism and user experience during simulator propulsion.

## Methods

### *Study Design Overview*

This study consisted of two sessions with 13 participants (10 able-bodied individuals and 3 experienced manual wheelchair users) separated by 3 days, in accordance with Jerald's (2016) recommendation of at least 2.5 days between VR exposures.

The first session was dedicated to simulator familiarization. The second session evaluated the ecological validity of the simulator, where participants performed propulsion tasks in both environments, first overground and then in the simulator, at self-selected speeds with instructions to match their overground performance. Biomechanical data were collected using instrumented wheels (IW) mounted on the wheelchairs, alongside participant feedback and perceived realism ratings.

### *Participants*

Participants were recruited regardless of gender or age, other than being major, through advertisements posted in the laboratory and on social media, with eligibility first assessed through phone interviews. All participants provided informed consent, as approved by the ethics committee of the Centre for Interdisciplinary Research in Rehabilitation of Greater Montreal (CRIR, Canada), ÉTS, and UQAM. Thirteen able-bodied individuals were recruited, of whom ten completed both sessions, along with three manual wheelchair users (MWUs) with more than six months of wheelchair use to ensure sufficient familiarity for evaluating simulator realism. Exclusion criteria were defined to ensure participants could safely and effectively complete the required tasks. These included an inability to perform selected basic propulsion skills listed in the Wheelchair Skills Test Questionnaire (WST-Q) Version 5.3 (Dalhousie University, 2022) as this would have excluded individuals with neurological or motor impairments preventing independent propulsion; shoulder or wrist pain greater than 6 on the Wheelchair User's Shoulder Pain Index (WUSPI), since individuals without major propulsion difficulties typically report values around 2 ± 2.5 (Curtis et al., 1995a, 1995b) and the presence of an ear infection or balance disorder, due to their potential impact on propulsion in virtual reality and association with motion sickness (Jerald, 2016). None of the participants were excluded based on these criteria.

Among able-bodied participants, the mean age was 27.1 years (SD = 6.1, range: 22–40), the mean height was 175.7 cm (SD = 9.4, range: 160–190), and the mean body weight was 75.0 kg (SD = 12.3, range: 57–95). They used a folding wheelchair provided by the laboratory, which weighed 14.1 kg with the instrumented wheels (IWs) installed. Eight participants were male, and two were female; eight were right-handed, and two were left-handed.

Three MWUs participated in the study. Their mean age was 49.3 years (SD = 15.5, range: 34–65), with an average height of 167.5 cm (SD = 14.8) and body weight of 64.2 kg (SD = 9.9). They had an average of 25.7 years of wheelchair experience (SD = 15.0, range: 11–41). MWUs used their own wheelchairs; after installation of the instrumented wheels, the mean wheelchair weight was 19.4 kg (SD = 1.3). Two participants were male, and one was female; two were right-handed, and one was left-handed.

### *Experimental Setup*

The protocol included two environments (overground and simulator), two task blocks (L-path and ramp), and multiple segments extracted from these blocks (acceleration, straight-line, turning, ascending, and descending), as presented in Table 1. Each task block in each environment began with a familiarization trial, followed by three experimental trials.

## *Overground Setup*

The L-shaped path was marked with cones to define segment separations, as shown in Figure 1a. The path consisted of a straight hallway followed by a 90° turn to the left (first and third trials) or to the right (second trial) to another hallway with the same width. The ramp is presented in Figure 1c and had a slope of 5.2°.

Biomechanical and kinematic data were collected using two instrumented wheels (modified SmartWheel (Asato et al., 1993; Chenier et al., 2021/2026)) mounted on the wheelchairs, and all trials were video-recorded for post-processing and precise segment separation.

**Table 1:** Experiment Sections: Environments: Overground and Simulator, Task Blocks: Path and Ramp, and Task Segments: Acceleration, Straight, Turning, Ascending, and Descending.

| **Environments** | **Overground** | | | | | **Simulator** | | | | |
|---|---|---|---|---|---|---|---|---|---|---|
| **Task Blocks** | **L-Path (3 trials)** | | | **Ramp (3 trials)** | | **L-Path (3 trials)** | | | **Ramp (3 trials)** | |
| **Task Segments** | Accel. | Str. | Turn | Ascend | Descend | Accel. | Str. | Turn | Ascend | Descend |

## *Simulator Setup*

The complete simulator structure has been described previously (Bayat & Chénier, 2025).

*Rolling Resistance Estimation.* To ensure that the simulated environment matched the overground environment, rolling resistance coefficients were calibrated in two steps. Coast-down tests were performed prior to participant testing, with a laboratory member completing three trials in the hallway (L-path) and three in the ramp area. Based on these results, coefficients of 0.01325 for the L-path and 0.024 for the ramp were applied. The inherited rolling resistance of the simulator ($\mu_sim = 0.0047$), as previously described (Bayat & Chénier, 2025), was subtracted from these values to finalize the coefficients used in the simulations.

*Virtual Environment and Control Integration.* Both environments were reproduced in Unity, as presented in Figure 1b and d. The slope was reproduced in real time by a D-Box platform that inclined the whole simulator, while the inertia and rolling resistance were reproduced by a haptic loop using Simulink running on a SpeedGoat system (Bayat & Chénier, 2025).

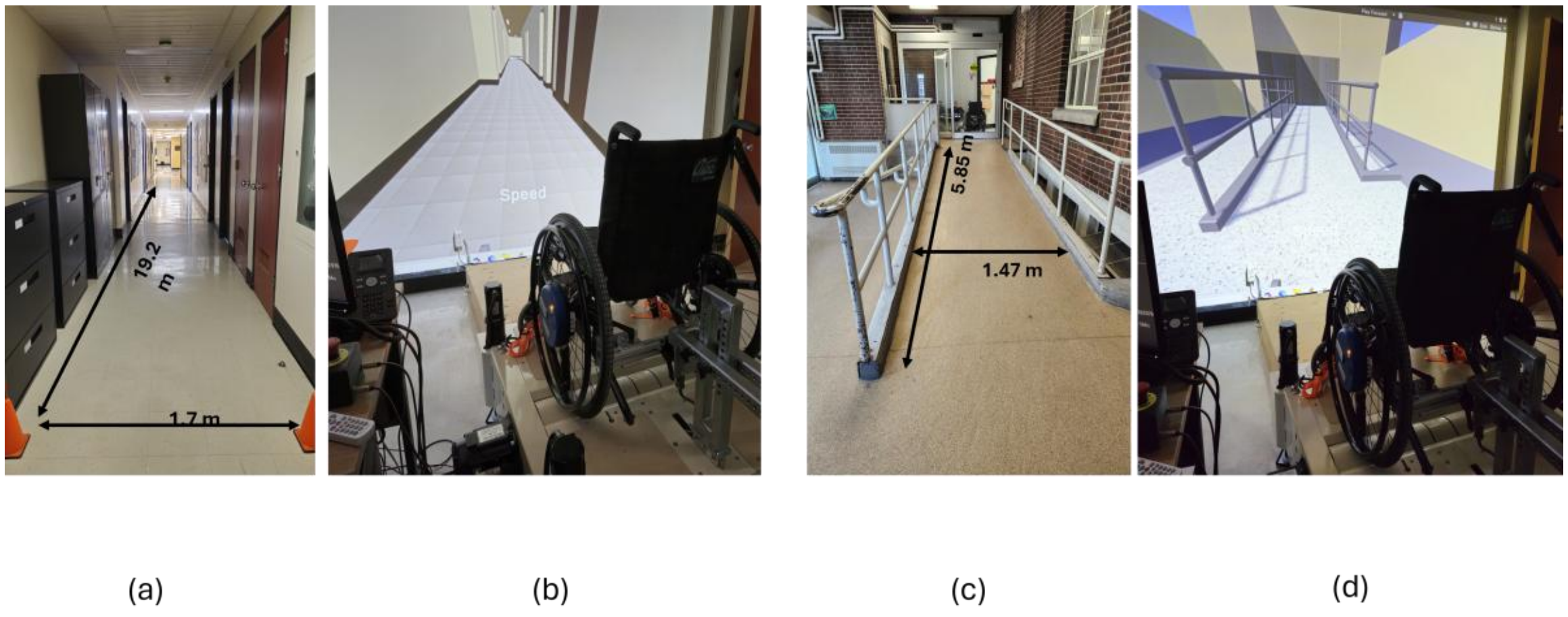

**Figure 1.** Overground and simulator scenes for the two task blocks. (a) Overground L-path straight-line segment; (b) simulator L-path straight-line segment; (c) overground ramp; (d) simulator ramp

## Experimental Protocol

### Day 1 – Familiarization

Given that the simulator was integrated with virtual reality (VR), precautions were taken to minimize the risk of motion sickness (MS). A dedicated familiarization session was implemented to provide gradual exposure to the simulator and thereby reduce the likelihood of MS (Jerald, 2016). Participants first propelled along a wide straight path, with exposure times progressively increased from 45 seconds to 1 minute and then 90 seconds. This progression was based on prior findings that 45 seconds of VR exposure while seated did not induce MS in most users (Riecke et al., 2015).

Subsequent familiarization involved ramp propulsion, beginning with a short ramp of mild slope (3.2°), followed by a medium ramp with a slope of 5°, and finally the target ramp, which replicated the overground configuration with a slope of 5.2° and identical length. Turning was practiced last, as it was considered the most likely to induce motion sickness. This progression began with a wide L-shaped turn, followed by a narrower turn, and concluded with the turn corresponding to the overground setup. Rest breaks were provided between stages, and participants were encouraged to report any discomfort or difficulties with the system.

### Day 2 – Validation

Testing began with task blocks in the overground environment, which were always performed first. After familiarization, participants completed three trials each on the L-shaped path and on the ramp. Participants completed the first trial at a comfortable self-selected speed, and the subsequent trials were performed while maintaining a similar speed. The mean speed across the three overground trials was recorded and later used as a target speed in the simulator. Ramp trials followed the same protocol, consisting of three ascending and descending segments, with the first performed at a comfortable speed and the next two conducted while maintaining that speed.

They then transitioned to the simulator, where the same sequence of tasks was performed in the corresponding virtual environments. For the L-path, participants matched their mean overground speed with the aid of real-time feedback presented on the screen, and three trials were conducted, with collisions resulting in trial repetition. For the ramp, participants propelled at a comfortable speed following familiarization and then performed three trials, with repetitions if collisions occurred. After each simulator task block (L-path and ramp), participants rated perceived realism on a 0–10 analog scale.

At the conclusion of the experimental tasks, participants completed a PIADS-inspired post-test questionnaire (Day & Jutai, 1996), adapted for this simulator. Finally, open-ended questions were conducted about what participants most liked and disliked, to gather qualitative feedback.

In addition to the main protocol, one of the expert MW users was asked to use the simulator with their own wheelchair without replacing the wheels to provide qualitative feedback on simulator realism. This exploratory assessment was conducted outside the formal experimental protocol.

## Quantitative Data Analysis

First, the dynamic offset from the instrumented wheel data was removed in accordance with the procedure described by Chénier et al. (2017). Push and recovery phases were then identified. Push onset was defined when the total force exceeded 5 N ($F_{tot} = \sqrt{F_x^2 + F_y^2 + F_z^2}$ ) .

The L-path Block trials were separated into three task segments: acceleration, straight, and turning. The exact start and end times of each segment were obtained manually from the videos for both overground and simulator environments.

Straight-line segments were defined by removing the first and last two pushes of each trial to isolate steady-state propulsion, while for acceleration, the first four strokes with the highest peaks in force and moment (Boninger et al., 2000) were analyzed. Turning segments were extracted from the L-path trials manually using video analysis, and ramp trials were divided into ascending and descending segments. Cycles from all three trials were combined for each segment and each hand, with turning data organized according to inward and outward hands. For task segments with clearly detectable pushes (straight-line, acceleration, and ascending), the ten most repeatable pushes—based on time-normalized Mz curves—were selected for each hand and participant, from which mean values of Mz, total force, power, velocity, cadence, and push angle were extracted for each environment. For turning and descending segments, where push detection was not possible or was unreliable due to task characteristics and variability in user technique, the entire segment duration was analyzed, and only total work and mean velocity were computed.

Finally, the instantaneous symmetry index (ISI) was calculated for each participant based on the moment about the wheel axle (Mz) from the ten most repeatable pushes in push-detectable segments, following the method described by Chénier et al. (2017). ISI values were low for all participants (OG: 0.13 ± 0.07; Sim: 0.25 ± 0.20), indicating minimal asymmetry between hands. Therefore, for push-detectable segments, variables from the dominant and non-dominant hands were averaged for subsequent analyses. However, total work was calculated as the sum of the work performed by both hands to represent the participant's total mechanical work.

To examine similarity between overground (OG) and simulator (Sim) environments, the straight-line, acceleration, and ascending segments were analyzed using a within-subjects multivariate analysis of variance (MANOVA) with a 2 × 3 design. The first independent variable was Environment (OG, Sim), and the second was Task Segment (straight-line, acceleration, ascending).

Nine dependent variables were analyzed: cadence, mean contact angle, mean push time, mean recovery time, mean moment about the wheel axle (Mz), mean total force (Fmean), mean power, mean velocity, and total work. Although speed was targeted during the straight-line simulator task block, actual mean velocity was retained as an outcome variable because speed matching was not imposed across all segments and because differences in achieved velocity reflect an important aspect of simulator realism.

Turning and descending segments were analyzed separately using a 2 × 2 repeated-measures MANOVA, with Environment (OG, Sim) and Task Segment (turning, descending) as within-subject factors. Total work was calculated as the sum of the work performed by both hands, and mean velocity was calculated over the full segment duration.

Effect sizes (Cohen's *d*) were reported as a follow-up analysis.

### *Propulsion pattern similarity*

Similarity between continuous curves for total force, power, and Mz was assessed using the Pearson correlation coefficient, based on the mean of the time-normalized ten most repeatable pushes for each participant. This approach provides a measure of waveform similarity independent of amplitude and offset differences. All statistical analyses were performed using SPSS (IBM Corp., Armonk, NY, USA).

## ***Qualitative Data Analysis***

Perceived realism ratings for the L-path and ramp tasks were analyzed quantitatively. Mean realism scores were calculated across all participants separately for each task block.

Questionnaire scores were reported separately for able-bodied participants and MWUs because the questionnaire items differed between groups. Items with negative wording, such as frustration and embarrassment, were reverse-scored before calculating the mean, so that higher scores consistently represented more favourable outcomes.

Audio recordings from the semi-structured interviews were automatically transcribed and, when necessary, translated from French into English using AI-assisted tools (OpenAI GPT-5.3 and Google Gemini 1.5 Pro). These tools also supported preliminary thematic extraction from open-ended responses, including reported likes, dislikes, and responses to selected questionnaire items. Coding criteria were defined by the researcher, who reviewed all transcripts and translations and conducted the final thematic coding. The most frequently reported themes were summarized descriptively.

## Results

### *Quantitative Data Analysis*

#### *Multivariate analysis*

A within-subjects MANOVA revealed no significant main effect of environment (Pillai's Trace = 0.097, $F(1,12) = 1.284$, $p = 0.279$, $\eta p^2 = 0.097$), indicating no overall difference between environments when averaged across task segments and variables. A significant main effect of task segment was observed (Pillai's Trace = 0.987, $F(2,11) = 414.496$, $p < 0.001$, $\eta p^2 = 0.987$), confirming substantial biomechanical differences between tasks. A significant environment × task interaction was found (Pillai's Trace = 0.874, $F(2,11) = 38.101$, $p < 0.001$, $\eta p^2 = 0.874$), indicating that differences between environments were task dependent. Additionally, a significant environment × task × variable interaction was observed ($p < 0.001$, $\eta p^2 = 0.780$; Greenhouse–Geisser corrected), suggesting that these task-dependent differences varied across biomechanical variables.

To further characterize these differences, effect sizes were calculated. The biomechanical outcomes for the straight-line, acceleration, and ascending segments are presented in Table 2 as mean, standard deviation and effect size d for both environments.

#### *Spatiotemporal parameters*

Although participants were instructed to match their overground speed on the simulator using the speed indicator, they generally propelled more slowly on the simulator during the straight-line and acceleration segments. Cadence was similar between environments during straight-line propulsion and acceleration, but participants used a lower cadence on the simulator during ascending.

Participants also used a smaller contact angle on the simulator during straight-line propulsion and acceleration, whereas the contact angle was similar between environments during ascending. Push time was shorter on the simulator during straight-line propulsion but longer during ascending. Recovery time showed non significant difference between environments, except during ascending, where it was higher on the simulator.

#### *Kinetic parameters*

Kinetic variables showed the clearest differences between environments. Participants generated lower mean Mz, total force, power, and work on the simulator than overground during the straight-line and acceleration segments. In contrast, ascending showed a different pattern: participants generated higher Mz and more total work on the simulator, while total force and power were more similar between environments.

Overall, the main pattern was that participants pushed less hard on the simulator during straight-line propulsion and acceleration, but showed equal or higher kinetic output during ascending, depending on the variable considered.

**Table 2:** Spatiotemporal and kinetic outcomes during straight-line, acceleration, and ascending segments. Values are mean (SD). *d* = Cohen's effect size; bold = medium effect, bold-underlined = large effect.

| | **Straight Line** | | | **Acceleration** | | | **Ascending** | | |
|---|---|---|---|---|---|---|---|---|---|
| | **OG** | **Sim** | **d** | **OG** | **Sim** | **d** | **OG** | **Sim** | **d** |
| Spatiotemporal variables | | | | | | | | | |
| Velocity (m/s) | 1.25 (0.23) | 1.14 (0.19) | **-0.79** | 0.96 (0.18) | 0.85 (0.15) | **-0.54** | 0.80 (0.20) | 0.73 (0.12) | -0.42 |
| Cadence (Pushes/min) | 52.7 (9.2) | 53.7 (9.3) | 0.15 | 47.3 (8.2) | 44.1 (7.4) | -0.45 | 59.1 (10.3) | 53.6 (8.0) | <u>**-0.85**</u> |
| Contact Angle (deg) | 27.3 (5.0) | 22.7 (4.7) | <u>**-1.48**</u> | 25.6 (4.8) | 22.8 (4.5) | **-0.54** | 21.0 (3.2) | 21.8 (3.3) | 0.18 |
| Push Time (s) | 0.41 (0.11) | 0.38 (0.09) | **-0.69** | 0.53 (0.09) | 0.57(0.12) | 0.38 | 0.51 (0.10) | 0.59 (0.11) | <u>**1.14**</u> |
| Recovery Time (s) | 0.77 (0.17) | 0.77 (0.22) | -0.02 | 0.72 (0.15) | 0.78 (0.17) | 0.49 | 0.52 (0.08) | 0.57 (0.13) | **0.54** |
| Kinetic variables | | | | | | | | | |
| Mz (Nm) | 6.5 (1.5) | 4.3 (1.6) | <u>**-2.68**</u> | 9.2 (2.2) | 6.9 (1.8) | <u>**-1.74**</u> | 23.0 (4.0) | 26.2 (4.2) | <u>**1.93**</u> |
| Total Force (N) | 39.1 (9.6) | 27.4 (8.6) | <u>**-2.46**</u> | 51.7 (13.9) | 37.7 (11.1) | <u>**-1.69**</u> | 113.2 (26.1) | 118.1 (24.1) | 0.43 |
| Total Power (W) | 25.9 (9.3) | 15.7 (6.6) | <u>**-2.41**</u> | 28.4 (9.4) | 18.5 (7.3) | <u>**-1.71**</u> | 60.4 (20.2) | 60.9 (18.8) | 0.05 |
| Total Work (J) | 537 (170) | 432 (188) | **-0.59** | 335 (76) | 213 (57) | <u>**-2.47**</u> | 1178 (190) | 1574 (310) | <u>**1.59**</u> |

For the turning and descending task segments, a multivariate analysis of variance (MANOVA) revealed significant main effects of task and variable, reflecting expected biomechanical differences between propulsion tasks and outcome measures. A significant environment × task × variable interaction was also observed, indicating that differences between simulator and overground environments were dependent on both the task and the biomechanical variable considered.

The biomechanical outcomes for the turning and descending tasks are presented in Table 3, where the entire duration of each trial was analyzed. In these tasks, no specific speed instruction was provided.

During turning, participants propelled more slowly on the simulator than overground, with a large effect. However, total work was nearly identical between environments. During descending, velocity was similar between environments, but total work was substantially more negative on the simulator, indicating greater braking effort.

**Table 3:** Spatiotemporal and kinetic outcomes during turning, and descending segments. Values are mean (SD). d = Cohen's effect size; bold = medium effect, bold-underlined = large effect.

| | **Turning** | | | **Descending** | | |
|---|---|---|---|---|---|---|
| | OG | Sim | d | OG | Sim | d |
| Velocity (m/s) | 1.08 (0.25) | 0.88 (0.12) | <u>**-1.06**</u> | 1.01 (0.19) | 1.06 (0.21) | 0.47 |
| Total Work (J) | 215 (88) | 215 (106) | 0 | -822 (127) | -1237 (123) | <u>**-4.01**</u> |

### *Propulsion pattern similarity*

Pearson correlation analyses showed statistically significant relationships between overground and simulator propulsion curves across all participants ($p < .001$). Correlations were generally high across total force, Mz, and power, with individual coefficients ranging from 0.82 to 0.97 in most cases.

Although the overall trend indicated strong similarity, variability was observed in specific cases. Participant 14 showed notably lower correlations during the acceleration segment (Force: 0.581; Mz: 0.413; Power: 0.537), and overall demonstrated the lowest agreement between environments ($r = 0.75 \pm 0.21$ across tasks and variables). Participant 13 also exhibited a reduced correlation for power during the ascending task (0.683).

In contrast, Participant 04 showed the highest level of similarity between overground and simulator environments, with an overall correlation of $r = 0.98 \pm 0.02$ across tasks and variables. Representative examples of the highest correlation (P04, Straight, Mz, $r = 0.99$) and the lowest observed correlation (P14, Acceleration, Mz, $r = 0.41$) are presented in Figure 2.

Among tasks, the straight-line segment yielded the most consistent results across participants and variables ($r = 0.95 \pm 0.04$). The ascending task showed slightly greater variability ($r = 0.92 \pm 0.08$), while the acceleration task exhibited the highest variability ($r = 0.93 \pm 0.12$). Group mean correlation values are summarized in Table 4.

**Table 4:** Pearson correlation coefficients between the simulator and overground force, Mz, and power waveforms

| | **Force** | **Mz** | **Power** |
|---|---|---|---|
| **Acceleration** | 0.949 | 0.930 | 0.930 |
| **Ascending** | 0.928 | 0.933 | 0.921 |
| **Straight** | 0.939 | 0.949 | 0.950 |

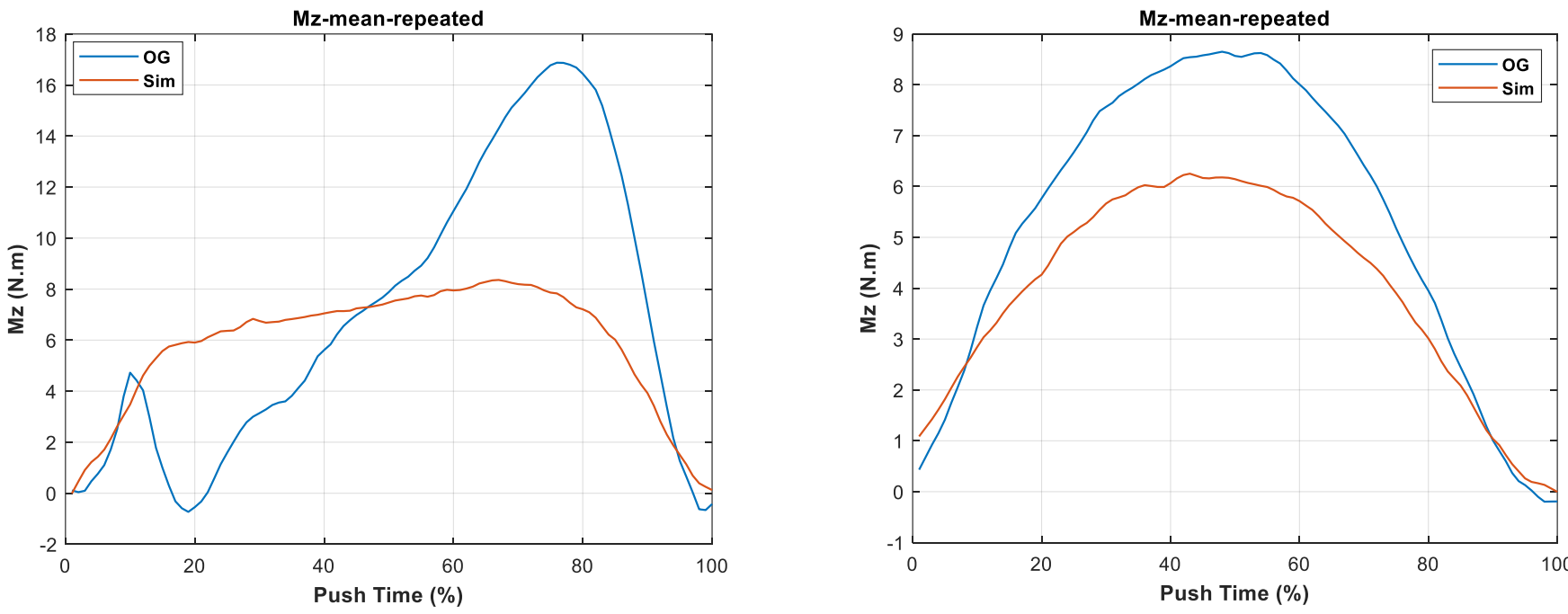

**Figure 2.** Representative propulsion waveforms showing (a) P14 – Acceleration – Mz (r = 0.41), and (b) P04 – Straight – Mz (r = 0.99)

## ***Qualitative Data Analysis***

Mean perceived realism ratings were $7.38 \pm 1.34$ for the L-path and $7.46 \pm 1.39$ for the ramp.

Post-task questionnaire results are summarized in Table 5, reported separately for able-bodied (A.B.) participants and manual wheelchair users (MWUs). The mean overall score was 9.02/10 for the able-bodied group and 6.30/10 for the MWUs group.

A summary of themes from open-ended participant feedback is listed below.

**Table 5:** Average adjusted questionnaire scores for manual wheelchair users and able-bodied participants.

| Question | MWUs Adjusted mean score | A.B Adjusted mean score |
|---|---|---|
| Ability to adapt to the activities of daily living | 2.00 | NA |
| Ability to participate | 5.00 | NA |
| Adequacy | 9.67 | 9.60 |
| Capability | 3.67 | 9.30 |
| Competence | 6.67 | 8.80 |
| Confusion | 6.00 | 7.40 |
| Eagerness to try new things | 7.33 | 8.90 |
| Efficiency | 8.33 | 9.10 |
| Embarrassment | 7.33 | 9.80 |
| Expertise | 2.00 | 9.30 |
| Frustration | 5.33 | 9.50 |
| Happiness | 7.00 | 8.90 |
| Independence | 7.00 | 9.10 |
| Performance | 9.00 | 8.40 |
| Productivity | 6.33 | 9.20 |
| Quality of life | 4.00 | NA |
| Security | 8.33 | 9.80 |
| Self-confidence | 9.33 | 9.20 |
| Self-esteem | 3.33 | NA |
| Sense of control | 5.33 | 8.00 |
| Well-being | 9.33 | NA |

### *Liked Most*

The most frequently mentioned liked theme was the ability to practice in a safe environment, allowing participants to take risks without fear. Eleven participants mentioned this during their interviews, using expressions such as being able to try dangerous actions like speeding down the ramp or performing maneuvers without the risk of colliding with others. This was especially emphasized in the ramp task block, which was perceived as riskier, particularly by able-bodied participants.

The second most liked theme was the realism of the simulation and physics, mentioned by six participants. They specifically referred to the ramp task block, including the feeling of gravity and tilt, as well as the realism of collisions.

Another liked theme was control learning and skill improvement, with participants highlighting that they could practice in a private environment without fear or danger from the real world. Able-bodied participants more explicitly stated that they felt their skills improved, whereas manual wheelchair users mentioned that the simulator has this potential, particularly during winter when outdoor propulsion is limited.

The final frequently liked theme was the fun and novelty of the experience. Participants mentioned the possibility of gamification and learning new techniques, and generally described the virtual experience as interesting.

### *Disliked Most*

The most frequently mentioned disliked theme was visual limitation, particularly a narrower field of view in the corridors. Participants reported difficulty estimating distance, which sometimes resulted in collisions. This was also associated with a lack of clear positional reference.

Another theme was steering sensitivity and control, with five participants stating that it was difficult to remain in a straight line. Some initially described propulsion as feeling harder on the simulator; however, after further questioning (e.g., whether this was due to resistance or control), it became clear that this perception was mainly related to difficulty maintaining a straight trajectory rather than increased mechanical resistance.

Regarding the ramp task block, many participants perceived descending as harder than overground, particularly because they felt they needed to brake more frequently.

Finally, only two able-bodied participants reported motion sickness, which they described as disturbing. In both cases, the test was paused to allow symptoms to subside. After a break and stepping out of the simulator, both participants were able to continue the experiment. Increasing the room lighting also appeared to help reduce symptoms during the remainder of the session.

The expert user, who tested the simulator using their own wheelchair without replacing the wheels, reported that overall, the experience felt more similar to real-life propulsion compared to using the IWs. They noted improved performance during ascending and better control when maintaining a straight-line task. The participant also highlighted the simulator's potential for training new wheelchair users, particularly by providing a controlled environment that may reduce stress and anxiety compared to real-world conditions. Additionally, they emphasized that the simulator allows users to safely practice real-life scenarios, potentially with the presence and support of family members.

## Discussion

The MANOVA results showed no significant overall difference between overground (OG) and simulator environments when all segments and variables were considered together. However, differences varied according to segment and variable, which is expected because wheelchair propulsion biomechanics are inherently task-dependent. Overall, spatiotemporal parameters appeared to be more preserved between environments, whereas kinetic parameters showed larger discrepancies.

The lower achieved speeds observed on the simulator are in line with participant feedback, as several participants mentioned that they did not perceive speed on the simulator as they would in real life. This may be related to the reduced visual flow of the non-immersive display. The lower speed observed in the present study is also consistent with previous studies reporting lower propulsion speeds on simulators compared with overground propulsion (Koontz et al., 2012; Lalumiere et al., 2014), although those studies focused only on straight-line propulsion. As suggested by Koontz et al. (2012), minor differences between the actual wheel speed and the speed presented on the screen, due to instrumentation errors, may also contribute to differences in achieved speed.

Other spatiotemporal findings are also consistent with previous studies. Lalumiere et al. (2014) reported a significantly lower contact angle on the LIO simulator, consistent with the straight-line results of the present study. Similarly, de Klerk, Velhorst, et al. (2020) reported lower cycle time and moderate differences in push time on the Esseda simulator when speed and power were matched. Additionally, Salimi & Ferguson-Pell (2019) reported a non-significant reduction in push length and an increase in cadence, aligning with the present findings. Regarding ramp propulsion, the only other study that compared OG and Sim environments in this condition found lower cadence during ascending (Klaesner et al., 2014), which is in line with the current study.

Pattern similarity analysis for force, moment, and power between overground (OG) and simulator environments showed excellent correlations, with r values greater than 0.92. This indicates that users tend to produce propulsion patterns similar to those observed during overground wheelchair propulsion.

These findings are consistent with previous studies reporting that participants tend to apply similar force patterns on simulators and during overground propulsion. Koontz et al. (2012) reported correlation values of approximately 0.9 for most participants, indicating strong similarity between force profiles in the two environments. Similarly, Lalumiere et al. (2014), using the LIO simulator, reported comparable propulsion patterns with correlation values around 0.9.

Overall, these findings indicate that propulsion patterns and temporal characteristics were relatively preserved on the simulator.

Kinetic variables showed the greatest differences between environments. Comparison with previous studies shows both agreement and divergence. Koontz et al. (2012) reported higher forces and torque on an ergometer despite lower speeds, which contrasts with the present findings. However, their study did not match speed or power between environments. Mason et al. (2014) also reported significant differences, particularly at higher speeds, where simulator limitations may become more pronounced.

Salimi & Ferguson-Pell (2019) evaluated a control-integrated simulator and reported comparable biomechanics between simulator and overground propulsion, with slightly higher, but non-significant, force values on the simulator.

De Klerk, Velhorst, et al. (2020) reported higher forces on the Esseda simulator when speed was matched, which they attributed to restricted body movement, while power and torque differences were minimal. In contrast, Lalumiere et al. (2014) reported lower power on the LIO simulator when participants matched perceived force rather than speed. Although effect sizes were not reported, the magnitude of this difference is comparable to that of the present study. Similarly, using the same simulator, Shaigetz et al. (2017) reported lower vertical force in the non-immersive VR condition.

These findings highlight that, although kinetic differences between simulator and overground propulsion are consistently reported, their direction is not uniform across studies. This variability is likely related to differences in simulator design, particularly in how rolling resistance and system dynamics are modelled. Koontz et al. (2007) demonstrated that even identical dynamometer systems can produce different results across sites due to variations in rolling resistance, calibration, and maintenance.

In the present study, the reduction in force and torque is consistent with findings reported for the LIO simulator, which shares similarities in its dynamic modelling approach. This suggests that modelling choices may influence the direction of observed kinetic differences.

In addition, in the studies discussed above, body movement was not included in the dynamic model equations or directly measured by the force instrumentation. Therefore, because the user's body movement did not affect wheelchair motion as it would in reality, participants may have tended to apply higher forces on the handrim, which could partly explain the higher kinetic values reported in those studies. In the present system, the force sensor measures forces applied to the wheelchair system, including those related to body movement, so this effect may have been partly accounted for. However, this measurement approach also appeared to be more sensitive than real overground propulsion, since even small body movements or minor asymmetries could influence simulator motion. This is consistent with participant comments indicating that wheelchair control was harder on the simulator, particularly in relation to steering and wheel sensitivity. Participants often described the wheels as moving with slight input, while also mentioning that they could maintain speed easily.

In the turning condition, no meaningful difference in total work was observed between environments, although high inter-participant variability was present. This was also reflected in participant feedback: some participants reported that turning felt very similar to reality, whereas others mentioned that the lack of peripheral view prevented them from anticipating turns and made control more difficult. Regarding comparison with other studies, although Salimi and Ferguson-Pell (2019) tested participants on the IAT path, they did not report straight-line and turning results separately. However, in the feedback they received from users, their second simulator model, which modelled rotational inertia using a pneumatic system, was the most valid and reliable biomechanically, but was the least preferred by participants, who felt that turning was not fast enough. In contrast, their third model, which modelled rotational inertia using only a coefficient, was the most favourable to participants, although its validity results were the weakest, making it the least biomechanically suitable. This suggests that perceived realism and biomechanical validity are not necessarily aligned.

From the participants' point of view, although perceived realism scores were high for the L-path, it was interesting that most participants perceived propulsion on the simulator as harder. Most clarified that this was mainly related to the control of the simulator being harder, while a few still felt that the force required to propel was higher on the simulator or that they experienced greater resistance. This may be explained by the most frequently reported limitations: the restricted field of view and lack of visual reference points, which made trajectory control more difficult and contributed to the perception that propulsion on the simulator was more challenging. This is consistent with findings by Shaigetz et al. (2017), who reported that non-immersive VR can make navigation more difficult because of its narrow field of view. Furthermore, the lack of proprioceptive and inertial cues in virtual environments may alter user behaviour (Salimi & Ferguson-Pell, 2019), which may help explain the observed differences in kinetic outcomes.

Kinetic differences were also evident during ramp propulsion. Klaesner et al. (2014) found higher force and Mz on the real ramp compared with the simulator, which is in the opposite direction of the present study. They explained that, in their simulator, the rollers did not turn backward, so users did not have to apply forces to compensate for backward rolling. In contrast, in our simulator, the rollers can rotate in both directions, and the force sensor measures the forces applied to the wheelchair system, including those related to gravity. The higher work observed during ascending and the more negative work observed during descending, indicating greater braking effort, are consistent with participant feedback describing ascending and descending as more demanding on the simulator. For example, some participants said that the ramp felt harder because they had to turn two wheels, referring to the wheelchair wheel and the roller. This also matches the feedback reported by Chaar and Archambault (2022) during testing of the MiWe simulator, where participants described ascending as harder and descending as harder to control, as if the resistance to movement changed. This similarity in user perception is interesting given the very different modelling of inertia and resistance in the two simulators. Despite this, almost all participants in the present study described the ramp as very realistic and true to life.

Perceived realism scores were slightly higher for the ramp task block than for the L-path. This trend was also reflected in responses to the open-ended questions, where participants described the ramp task block as more realistic, particularly due to the sensation of gravity and tilt. The ramp was also the task block in which participants most frequently mentioned feeling safer and more confident, as it helped them become more familiar with real-life scenarios.

These findings are consistent with previous studies reporting improved presence and reduced motion sickness in wheelchair simulation contexts when motion feedback is added. For example, Vailland et al. (2020) reported enhanced presence and lower motion sickness in a power wheelchair simulator, while Saito et al. (2025) demonstrated the positive effect of motion feedback in reducing anxiety during ramp navigation.

The themes of safety and the ability to practice without fear are particularly important, as they may support better skill acquisition. Shaigetz et al. (2017) noted that wheelchair training often occurs in constrained clinical environments, which may limit skill transfer, while time constraints further restrict practice opportunities. In this context, this simulator may provide a valuable complementary training approach, particularly for anxious users (Lettre et al., 2023). Additionally, the ability to train indoors may be beneficial in challenging weather conditions.

Only 15% of participants (two individuals) experienced motion sickness, both of whom reported prior susceptibility. Although this rate is slightly higher than in previous studies (Arlati et al., 2020; Chaar &

Archambault, 2022), symptoms were mainly associated with curvilinear motion, which was not included in those studies.

In general, manual wheelchair users reported lower questionnaire scores compared to able-bodied participants. However, all MW participants explained that lower ratings for items such as perceived capability reflected their existing skills rather than dissatisfaction with the simulator, and they indicated that novice users would likely benefit more. A similar explanation was provided for items related to adaptation, expertise, and self-esteem. Despite these lower scores, overall scores remained high (9/10 for able-bodied participants and 6.3 for MW users) and increased to approximately 7 when excluding these specific items.

The simulator performed comparably to existing systems, with kinetic discrepancies consistent with those reported in the literature and good agreement in temporal parameters. In particular, waveform similarity for force, Mz, and power supports its ability to reproduce realistic propulsion patterns. Overall, the simulator showed a level of agreement with overground propulsion that was within the range reported for other wheelchair simulators, while showing similar or smaller discrepancies in some temporal variables, such as cadence and cycle time.

A key advantage of this system is its ability to be used with the user's own wheelchair and wheels, preserving ergonomics and probably facilitating clinical usage. This was supported by exploratory feedback from one of the expert users, who reported improved perception of realism when using their own wheelchair and wheel. In addition, the integration of multiple sensory feedback modalities, including the motion platform, may contribute to a greater perceived realism, although transfer to real-world propulsion should be examined in future work. However, based on participant feedback, several improvements could further enhance realism, including refining virtual scenes, incorporating gamification, adding visual reference cues, and integrating additional sensory inputs such as sound and airflow. Participants also suggested that including peripheral visual input could improve performance during turning.

The use of a force sensor as the primary measurement instrument, instead of instrumented wheels, provides the advantage of allowing users to operate their own wheelchairs without modification. However, this approach introduces limitations, including increased sensitivity to positional drift, which may explain the slight drift reported by some participants.

The sample size was limited, particularly with respect to experienced manual wheelchair users. Including novice users instead of able-bodied participants would provide more representative results, but this was not feasible in the present study. In addition, the use of instrumented wheels for data collection may have influenced perceived realism, as was clearly reflected in the exploratory feedback of one of the expert users.

Future work should address these limitations by improving the control system, as well as by recruiting a larger and more diverse participant population, including experienced users and patient partners.

Given the agreement observed in temporal parameters, the use of this simulator for providing biofeedback without relying on instrumented wheels can be evaluated for future research.

## Conclusion

In this study, the ecological validity of the proposed simulator was evaluated by comparing biomechanical outcomes between environments and by analyzing participant feedback on perceived realism. The results showed good agreement in temporal parameters, with mostly negligible to small differences across tasks. In contrast, kinetic variables exhibited larger discrepancies between environments, generally with lower values in the simulator. Total work was lower in the simulator for most tasks, except during ascending and descending, where higher magnitudes were observed.

Questionnaire and interview results indicated a positive perception of the simulator. Participants generally reported high realism scores. The most frequently reported positive themes were the ability to practice in a safe and controlled environment and to explore new skills without risk. The main limitations identified were the restricted field of view and the lack of visual reference cues, which made trajectory control more challenging.

Overall, the simulator demonstrates a level of discrepancy in kinetic variables comparable to that reported for other simulators, while achieving similar or improved agreement in temporal parameters and waveform similarity. A key advantage of this system is the ability to use the user's own wheelchair and wheels, which could preserve ergonomics and improve user comfort, thereby supporting potential clinical applications.

Future work will focus on evaluating the capability of the simulator to provide real-time biofeedback to users.

## Acknowledgements

The authors would like to thank the participants for their time and valuable feedback, and Nicolas Fleury-Rousseau for his assistance during data collection.

## Disclosure statement

No potential conflict of interest was reported by the authors.

## Funding

This work was supported by the Fonds de recherche du Québec – Nature et technologies and the BRILLIANT-Rehab program via the Canada Foundation for Innovation.

## Ethics approval

This study received ethics approval through the Centre for Interdisciplinary Research in Rehabilitation of Greater Montreal (CRIR) from the Research Ethics Committee in Rehabilitation and Physical Disability (CER RDP) of the CIUSSS du Centre-Sud-de-l'Île-de-Montréal (project no. 2024-1792). This approval was subsequently recognized by the Research Ethics Committee of the École de technologie supérieure (ÉTS; reference H20240503) and by the Research Ethics Committee involving Human Participants (CIEREH) of the Université du Québec à Montréal (UQAM; certificate no. 2025-7151). All participants provided written informed consent before participation.

## Data availability statement

The data that support the findings of this study are available from the corresponding author upon reasonable request, subject to ethics approval and participant confidentiality restrictions.

## Declaration of generative AI use

ChatGPT (OpenAI, GPT-5.5 Thinking) was used for language refinement and occasional clarification or summarization of author-provided text. OpenAI GPT-5.3 and Google Gemini 1.5 Pro were also used for transcription, translation, and preliminary thematic extraction, as described in the Methods. The authors reviewed and verified all AI-assisted content and analyses and remain responsible for the manuscript.